\documentclass[openacc]{rstransa}
\usepackage[amssymb]{SIunits}
\usepackage[square,numbers,sort&compress]{natbib}
\usepackage[normalem]{ulem}

\usepackage{color}

\begin{document}

\title{Sodium chloride inhibits effective bubbly drag reduction in turbulent bubbly Taylor--Couette flows}

\author{
Luuk J. Blaauw$^{1}$, Detlef Lohse$^{1,2}$ and Sander G. Huisman$^{1}$}

\address{$^{1}$Physics of Fluids Group and Max Planck Center for Complex Fluid Dynamics, Department of Science and Technology, and J.M. Burgers Center for Fluid Dynamics, University of Twente, Enschede, The Netherlands\\
$^{2}$Max Planck Institute for Dynamics and Self-Organization, Am Fa\ss berg 17, G\"ottingen, Germany}

\subject{Physics, Fluid Mechanics}


\corres{Sander G. Huisman\\
\email{s.g.huisman@utwente.nl}}

\begin{abstract}
Using the Taylor--Couette geometry we experimentally investigate the effect of salt on drag reduction caused by bubbles present in the flow. We combine torque measurements with optical high-speed imaging to relate the bubble size to the drag experienced by the flow. Previous findings have shown that a small percentage of air (4\%) can lead to dramatic drag reduction (40\%). In contrast to previous laboratory experiments, which mainly used fresh water, we will vary the salinity from fresh water to the average salinity of ocean water. We find that the drag reduction is increasingly more inhibited for increasing salt concentration; going from 40\% for fresh water to just 15\% for sea water. Salts present in the working fluid inhibit coalescence events, resulting in smaller bubbles in the flow and with that, decreasing the drag reduction. Above a critical salinity, increasing the salinity has no further effect on the bubble sizes in the flow and thus the drag experienced by the flow. Our new findings demonstrate the importance of NaCl on the bubbly drag reduction mechanism, and will further challenge the naval architects to implement promising air lubrication systems on marine vessels.
\end{abstract}


\begin{fmtext}
\end{fmtext}
\maketitle
\section{Introduction}\label{sec:Intro}
Most goods are transported over sea by large container ships. The largest contribution of the drag force of these ships is caused by the skin friction on the ship hull \cite{larsson_2010_ship,Lackenby_1962_ship}. Reducing the skin friction can lead to substantial decrease in the ship's fuel consumption. Not only is this financially attractive, but also it reduces the emissions of $\text{CO}_2$ substantially. One promising technique for reducing the skin friction is by injecting air into the flow under the ship's hull. Other advantages of this system are that it does not pollute the ocean, nor disturb marine fauna.\\
Drag reduction by air lubrication can be achieved in multiple ways. Either by air layers \cite{elbing_2008,elbing_2013,Verschoof_2018_cavities,Zverkhovskyi_2014}, reducing the contact area of water with the ship hull, or by bubbles altering the turbulent flow in the boundary layer below the hull. We will focus on the latter method, i.e., on bubbly drag reduction, which has been reviewed by refs.\ \cite{Ceccio_2010_review,Murai_2014_review,Hashim_2015_review,lohse_2018}. In those review papers multiple possible mechanisms for bubbly drag reduction are treated. In flat plate experiments up to 80\% of drag reduction has been observed using small bubbles at a Reynolds number of $\text{Re}\approx2\times10^6$ \cite{Madavan_1984,Madavan_1985}. Also in numerical simulations of channel flow \cite{xu_2002} it was found that non-deformable micro bubbles can reduce the drag up to $6.2\%$ for a moderate Reynolds number of $\text{Re}=3000$. This suggest that the bubble size perhaps does not play an important role in the amount of drag reduction, suggesting that the void fraction is the determining factor \cite{Ceccio_2010_review,Moriguchi_2002}. Compressibility (e.g. ref.\ \cite{Lo_2006}) and turbulent suppression (e.g. refs.\ \cite{kato_1999,ferrante_elghobashi_2004}) are also considered to be possible mechanisms for bubbly drag reduction. Furthermore, deformability of bubbles can play a major role in the effects of bubbly drag reduction \cite{vdBerg_2005,vGils_2013_bubbledeform,Kitagawa_2005_bubbledeform,Lu_2005_bubbledeform,Verschoof_2016_surfactant}.\\
Different drag-reducing mechanisms apply for low and high Reynolds number Taylor--Couette flows. In low Reynolds number flows small bubbles rise due to gravity and destroy the Taylor vortices, removing a very efficient transport mechanism of angular momentum between the cylinders \cite{Murai_2005,murai_2008,sugiyama_2008}. For high Reynolds number Taylor--Couette flows it was found by high-performance direct numerical simulations \cite{spandan_2018} that the deformability of bubbles plays an important role in the drag reducing mechanism.\\
One open question is how bubbly drag reduction can be applied to marine vessels. A few ships have already been equipped with bubble injection systems. Examples are the Silverstream project \cite{silverstream_site} and the Mitshubishi Air Lubrication System (MALS) \cite{Mizokami_2010_mitsubishi,Mizokami_2013_mitsubishi}. The last one achieved around 10\% of power savings, already taking the energy needed for the air injection into account. Similar power savings were achieved using a hydrofoil on a variety of ships \cite{Kumagai_2015}. These power savings are however not nearly as large as the drag reductions measured in the lab. A possible explanation for the discrepancy between laboratory measurements and measurements on marine vessels could be size effects when scaling the flow from lab scale to ships \cite{Ceccio_2010_review,Kodama_2000,Jang_2014}, surface details of the hull like roughness \cite{vdBerg_2007,Bullee_2020_hydrophobic,Bullee_2020_roughness, Deutsch_2004,vanBuren_2017}, mineral/salt content of the water, particulates content, or surfactants \cite{Verschoof_2016_surfactant}.\\
Almost all lab studies are performed with tap water or even cleaner water \cite{vanBuren_2017,sugiyama_2008,vdBerg_2005,vGils_2013_bubbledeform,ezeta_2019,spandan_2018,gillissen_2013,sanders_2006, Madavan_1984,Madavan_1985,murai_2007}. Seawater, however, contains a large variety of chemicals and particles. Few lab studies have been performed with `contaminated' water. For example, Winkel \emph{et al}.\ \cite{Winkel_2004} measured bubble size distributions in salt water in a horizontal water channel and found that the bubble sizes decrease monotonically for increasing salt concentrations. In contrast, Shen \emph{et al}.\ \cite{shen_2006} found that by injecting bubbles into solutions with salts or surfactants that the drag reduction is independent on the bubble size. Furthermore, Elbing \emph{et al}. \cite{elbing_2008} measured the drag reduction in a surfactant solution in a water channel and found that the resulting size of the bubbles downstream of the injection and the drag reduction do not significantly depend on the surfactant concentration. In contrast, Verschoof \emph{et al}. \cite{Verschoof_2016_surfactant} found that the injection of only 6 ppm of surfactant had a dramatic effect on the drag reduction in a Taylor--Couette setup, reducing it from 40\% to just 4\%, for a global air volume fraction of $\alpha= 4\%$. They estimated the Weber number $\text{We} = {\rho u'^2 d_{\text{bubble}}}/{\gamma}$, where $\rho$ is the fluid density, $u'$ are the velocity fluctuations, $d_\text{bubble}$ the bubble diameter, and $\gamma$ the surface tension. The Weber number quantifies the deformability of the bubbles, by comparing the drag force of a bubble with surface tension forces. Verschoof \emph{et al}. \cite{Verschoof_2016_surfactant} found that bubbles with higher Weber numbers, i.e., deformable bubbles, are much more effective for drag reduction. This was in accordance with earlier observations \cite{vGils_2013_bubbledeform,Kitagawa_2005_bubbledeform}. It is important to note that there are certain differences between a water channel and a Taylor--Couette setup. Whereas in channel flow and flat plate experiments there is a distinct gravitational force, in Taylor--Couette flow bubbles experience a centripetal force (and the effect of gravity is minimal for the fully turbulent case). Also, in Taylor--Couette flow there is no developing boundary layer such as is the case in channel flow or flat plate experiments.\\
Salts have a marked effect on the kinetics and dynamics of bubbles in flows \cite{Craig_1993,Hori_2020}. Bubble sizes in turbulent flows are governed by a dynamic equilibrium between bubble splitting and bubble coalescence events. This equilibrium can be affected by salts which inhibits the coalescence of bubbles. Some combinations of ions will inhibit the coalescence of bubbles, whereas others do not have an effect on bubble coalescence \cite{Craig_1993}. Sodium chloride is among those salts that inhibit bubble coalescence. With increasing salinity the equilibrium bubble size decreases monotonically, indicating that the inhibition of coalescence becomes stronger. At a critical concentration the equilibrium bubble size is not affected anymore \cite{Firouzi_2014,Firouzi_2015,Quinn_2014,Sovechles_2015}. This critical concentration depends on the type of salt and on the bubble size \cite{Tsang_2004,Firouzi_2015}.\\
In this paper we investigate the effect of water quality, more specific that of the sodium chloride concentration, on bubbly drag reduction. We employ the Twente Turbulent Taylor--Couette (T$^3$C) facility \cite{vGils_2011_T3C}, with which we have before measured bubbly drag reduction in clean water \cite{vGils_2013_bubbledeform}. The concentration of sodium chloride is varied between decalcified water to the concentrations found in ocean water. We measure the friction of the flow on a wall. The size of the bubbles is measured using high speed imaging, and we relate this to the measured drag reduction.\\
The paper is built up in the following way. In section \ref{sec:Exp_Setup} we introduce the T$^3$C facility, the setup, and the experimental conditions. Section \ref{sec:Results} discusses the results. We conclude our research in section \ref{sec:Conclusions}.\\

\section{Experimental Setup \& Control Parameters}\label{sec:Exp_Setup}
\begin{figure}
    \centering
    \includegraphics[width=0.5\textwidth]{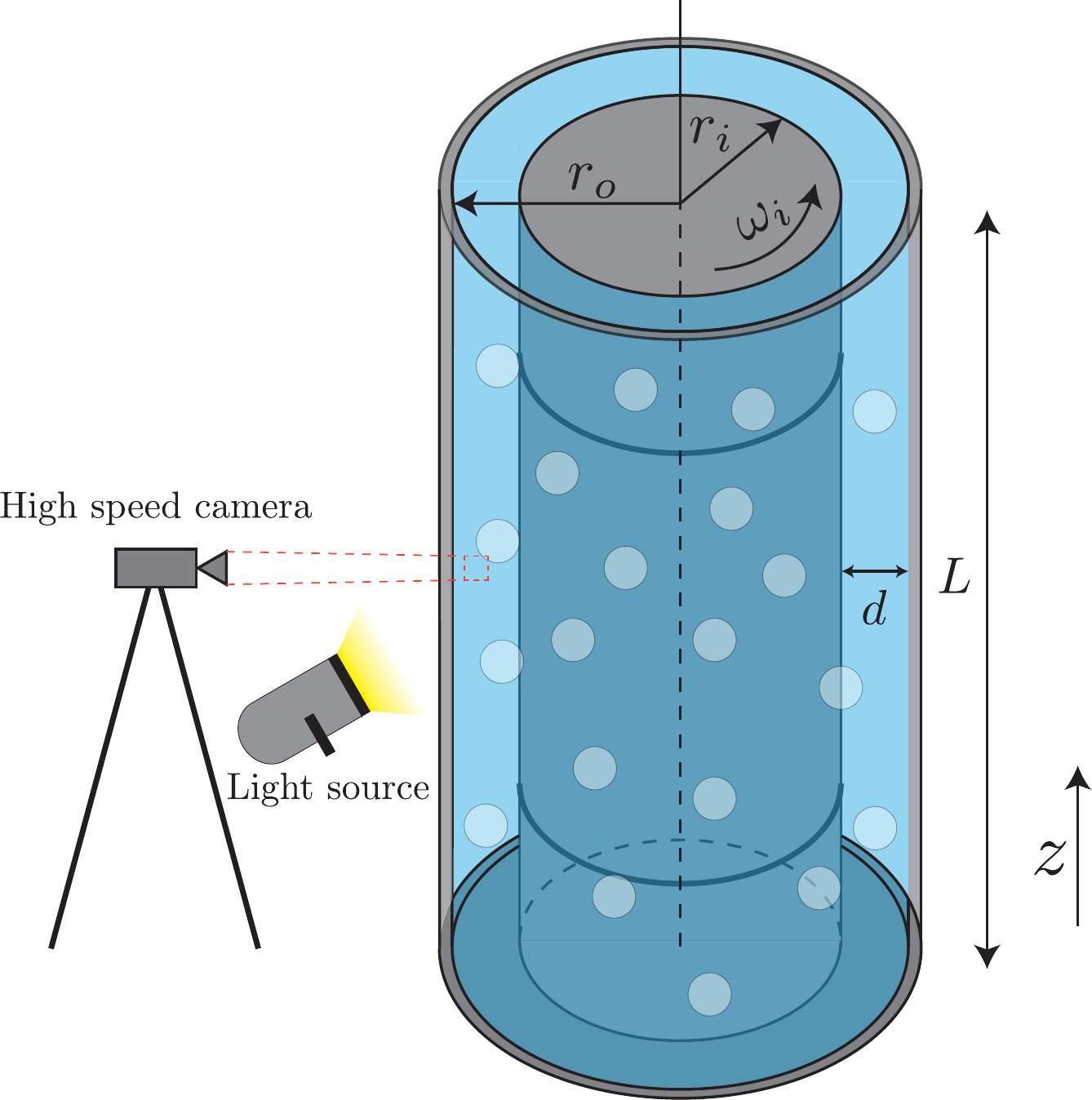}
    \caption{Schematic of the Twente Turbulent Taylor--Couette (T$^3$C) setup \cite{vGils_2011_T3C} (bubbles not to scale). Solid lines on the inner cylinder show the three sections of the inner cylinder. Torque is measured on the middle section. Red dashed lines show the field of view (\unit{30}{\milli\meter}$\times$\unit{30}{\milli\meter}) of the camera. Camera is positioned at midheight, $z=L/2$.}
    \label{fig:ExpS_ExpSetup}
\end{figure}

The Taylor--Couette geometry \cite{Taylor_1923} consists of two concentric cylinders that can rotate independently. Taylor--Couette flow displays a wide variety of flow phenomena \cite{andereck_1986,Grossmann_2016_TCrev,ostillamonico_2014}. The system has been referred to as the hydrogen atom of fluid mechanics \cite{Fardin_2014}. The Taylor--Couette geometry is a closed system and has therefore a well-defined energy balance, where all the energy inserted by driving the flow is dissipated by the turbulence \cite{Eckhardt_2007}.\\
In analogy with Rayleigh--B\'enard convection (the flow between a heated bottom plate and a cooled top plate) where heat is transported \cite{ahlers_2009}, in Taylor--Couette flow, angular velocity is transported \cite{Eckhardt_2007}. In Rayleigh--B\'enard convection the heat transport is characterized with the Nusselt number (the convective heat transport over the conductive heat transport) and in analogy, in Taylor--Couette flow, the angular velocity transport can be characterized by the angular velocity Nusselt number, $\text{Nu}_\omega$, which is the ratio of the angular velocity transport over the angular velocity transport in the laminar, non-vortical, flow state. This Nusselt number can be expressed as $\text{Nu}_\omega = \tau/\tau_\text{lam}$, where $\tau$ is the torque measured on the inner cylinder and $\tau_\text{lam}=4\pi L\nu_l\rho_l ( \omega_i-\omega_o ) (r_i^2r_o^2)/(r_o^2-r_i^2)$ is the torque on the inner cylinder for the laminar (non-vortical) case, with $L$ the length of the inner cylinder, $\nu_l$ and $\rho_l$ the viscosity and density of the fluid, respectively, $\omega$ the angular velocity, and $r$ the radius of the cylinder, where the subscripts $i$ and $o$ refer to the inner and outer cylinder, respectively.\\
Whereas in Rayleigh--B\'enard convection the (thermal) driving is given by the Rayleigh number, in Taylor--Couette flow the (mechanical) driving is given by the Taylor number: 
\begin{align}
    \begin{split}
        \text{Ta} = \frac{1}{4}\sigma\frac{(r_o-r_i)^2(r_o+r_i)^2(\omega_i-\omega_o)^2}{\nu^2}
    \end{split}
\end{align}
where $\nu$ is the kinematic viscosity of the working fluid, and $\sigma = (1+\eta)^4/{16\eta^2}$ can be seen as the quasi-Prandtl number \cite{Eckhardt_2007}, solely determined by the radius ratio $\eta=r_i/r_o$. For measurements where the working fluid contains bubbles we correct the density and the viscosity using the Einstein equation \cite{einstein_1905} as was done previously \cite{vGils_2013_bubbledeform,Bullee_2020_hydrophobic,Bullee_2020_roughness,Verschoof_2016_surfactant,ezeta_2019}:
\begin{align}
    \rho &= \left(1-\alpha\right)\rho_l \label{eq:einstein_rho} \\
    \nu &= \left(1+ \frac 52 \alpha \right) \nu_l \label{eq:einstein_nu}
\end{align}
where $\rho_l$ and $\nu_l$ are the density and the viscosity of the carrier liquid (see table \ref{tab:fluidprop}), respectively, and $\alpha$ is the global air volume fraction.\\
The experiments are performed in the Twente Turbulent Taylor--Couette (T$^3$C) setup \cite{vGils_2011_T3C}, see figure \ref{fig:ExpS_ExpSetup} for a schematic. The acrylic outer cylinder allows for good optical access to the flow. Multi-phase flows have been widely studied in the Taylor--Couette geometry \cite{Chouippe_2014,Climent_2007,Djeridi_2004,Mehel_2005,bakhuis_2021} often in light of drag reduction by air lubrication \cite{Bullee_2020_hydrophobic,Bullee_2020_roughness,vdBerg_2005,vdBerg_2007,Verschoof_2016_surfactant,vGils_2013_bubbledeform,Fokoua_2015,Murai_2005,sugiyama_2008,murai_2008}\\
In the current study only the inner cylinder of the T$^3$C is rotated, while the outer cylinder is kept stationary. The inner and outer cylinder have a radius of $r_i = \unit{0.2000}{\meter}$ and $r_o = \unit{0.2794}{\meter}$, respectively, resulting in a gap width of $d=r_o-r_i=\unit{0.0794}{\meter}$ and a radius ratio of $\eta={r_i}/{r_o} = 0.716$. The height of the setup is $L = \unit{0.927}{\meter}$, giving an aspect ratio of $\Gamma=L/d= 11.7$. \\
As indicated in figure \ref{fig:ExpS_ExpSetup} the inner cylinder is divided into three sections. We only measure the torque on the middle section of the inner cylinder to prevent any influence of the (stationary) end plates. We connect the driveshaft with the middle inner cylinder using an Althen hollow axis torque transducer with a range of $\pm$ \unit{225}{\newton\meter} and $0.25\%$ accuracy. For more details on the T$^3$C setup, see ref.\ \cite{vGils_2011_T3C}.\\
The measurement volume is cooled by means of actively cooling the bottom and top end plates. Temperature inside the measurement volume is measured with three Pt100 temperature sensors mounted flush in the inner cylinder. During a measurement the temperature of the working fluid is kept constant within \unit{1}{\kelvin}. The viscosity and density of the working fluid are calculated using the measured mean temperature and their values at $T=\unit{21}{\degreecelsius}$ are listed in table \ref{tab:fluidprop}.\\
As working fluid we use decalcified water. The water was analyzed using ion chromatography to measure the concentrations of ions already present in the working fluid, see table \ref{tab:Exps_Ions}. The water contains little calcium and magnesium, but does contain sodium and chloride. However, we found these concentrations to be negligible compared to the concentrations used in our experiments. Initial experiments by us using ultrapure MilliQ water have shown no different results from those with decalcified water.\\
\begin{table}
    \centering
    \caption{Concentration of different cations and anions present in our working fluid (decalcified water).}
    \begin{tabular}{|c|c|}\hline
        Ion & Concentration\\
        & $[\unit{}{\milli\gram\per\kilogram}]$ \\
        \hline
        $\text{Na}^+$ & 86.27\\
        $\text{K}^+$ & 3.23\\
        $\text{Ca}^{2+}$ & 0.69\\
        $\text{Mg}^{2+}$ & 0.052\\\hline
        $\text{Cl}^-$ & 19.84\\
        $\text{NO}_3^-$ & 2.63\\
        $\text{SO}_4^{2-}$ & 45.48\\\hline
    \end{tabular}
    \label{tab:Exps_Ions}
\end{table}
The measurement volume is filled entirely with water for the reference case. In the drag reduction cases we fill the volume for 96\% with water and leave 4\% filled with air, i.e., the average air volume fraction is $\alpha = 4\%$. The turbulence is strong enough to entrain all the air and distribute the bubbles throughout the entire domain. We add industrial grade sodium chloride to the system in concentrations up to \unit{35}{\gram} NaCl{\per\kilogram} solution ($3.5\% (w/w)$), corresponding to the concentration of salt present in standard ocean water. We use the relative salinity, defined in equation \ref{eq:ExpS_salinity}, where $S$ is the salinity in the measurement and $S_\text{ocean}$ is a salinity of \unit{35}{\gram} of salt per kilogram of solution.\\
\begin{equation}\label{eq:ExpS_salinity}
    \Tilde{S} = \frac{S}{S_\text{ocean}}
\end{equation}
We utilize high speed photography, simultaneous to our torque measurements, to measure the bubble size in the flow, see figure \ref{fig:ExpS_ExpSetup}. We use a Photron Mini AX200 high speed camera with a \unit{300}{\milli\meter} objective to get a field of view of approximately $\unit{30}{\milli\meter}\times\unit{30}{\milli\meter}$. We capture images at \unit{50}{fp\second} such as to have uncorrelated images to obtain the bubble size distribution. The camera is focused close to the outer cylinder, $r_\text{focus}\approx\unit{0.275}{\meter}$, as the bubbles prevent direct optical access to bubbles close to the inner cylinder. We do not expect a large radial dependence on the bubbles size due to the strong radial mixing \cite{Huisman_2012}, see also figure 10 of van Gils \emph{et al.}\ \cite{vGils_2013_bubbledeform} where $d_\text{bubble}(r)$ is measured for the Taylor--Couette geometry.\\

\begin{table}[h]
\centering
\caption{Material properties for our working liquid at $T=\unit{21}{\degreecelsius}$: $\rho_l$ and $\nu_l$ are from refs.\ \cite{Pereira_2001_dens} and ref.\ \cite{Aleksandrov_2012_visc}, respectively, $\gamma$ is taken from the model of ref.\ \cite{Dutcher_2010_surfacetension}, and $\rho$ and $\nu$ (so including the effects of bubbles) are calculated according to equations \ref{eq:einstein_rho} and \ref{eq:einstein_nu}, respectively, using $\alpha=0.04$.}
\begin{tabular}{|c|c|c|c|c|c|c|}
\hline
    $\tilde{S}$ & $\rho_l$ &  $\nu_l$ & $\gamma$  & $\rho$ & $\nu$ & $\rho/\gamma$ \\
    & $[\unit{}{\kilogram\per\meter^3}]$ & $\times10^{-6}[\unit{}{\meter^2\per\second}]$ & [$\unit{}{\milli\newton\per\meter}]$ & $[\unit{}{\kilogram\per\meter^3}]$ & $\times10^{-6}[\unit{}{\meter^2\per\second}]$ &$\times10^{4}[\unit{}{\second^2\per\meter^3}]$\\
    \hline
    0 & 998.1 & 0.9778 & 72.8 & 958.2 & 1.0756 & 1.317\\
    0.5 & 1011.1 & 1.0034 & 73.2 & 970.6 & 1.1038 & 1.325\\
    0.75 & 1017.6 & 1.0163 & 73.5 & 976.9 & 1.1179 & 1.330\\
    1 & 1024.1 & 1.0299 & 73.7 & 983.1 & 1.1329 & 1.334\\
    \hline
\end{tabular}
\label{tab:fluidprop}
\end{table}

\section{Results and Discussion} \label{sec:Results}
The torque on the inner cylinder is directly related to the skin friction on the inner cylinder, which is of key interest for marine applications. Out of the measured torque we calculate the drag. Using high speed imaging we measure the sizes of the bubbles in the flow, which is used to estimate the Weber number, characterizing the deformability of the bubbles. We relate the deformability to the measured drag reduction, since deformability has proven crucial for effective drag reduction \cite{Kitagawa_2005_bubbledeform,Lu_2005_bubbledeform,vGils_2013_bubbledeform,Verschoof_2016_surfactant, lohse_2018}.

\subsection{Torque measurements} \label{sec:Results_torque}
\begin{figure}
    \centering
    \includegraphics[width=0.49\textwidth]{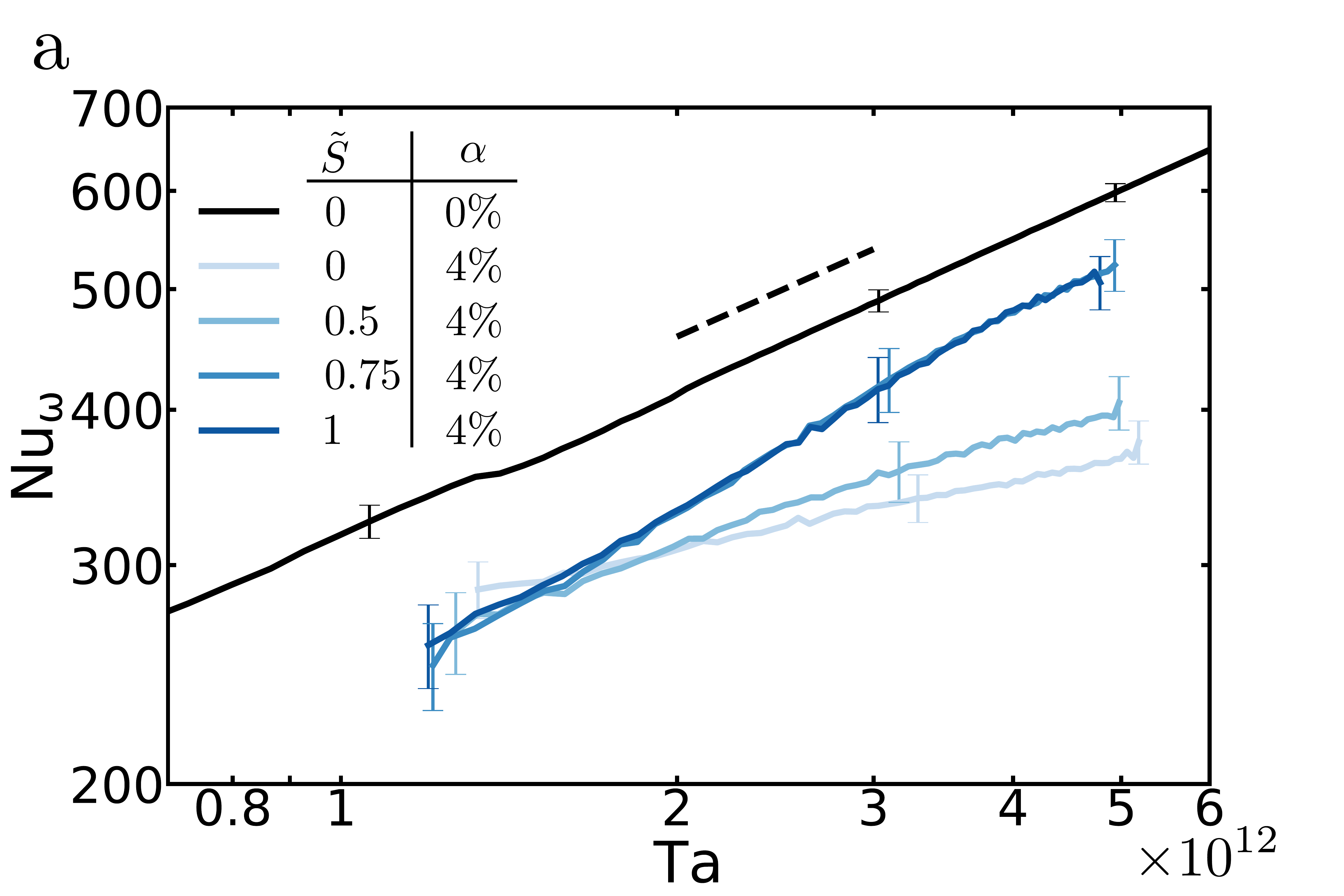}
    \includegraphics[width=0.49\textwidth]{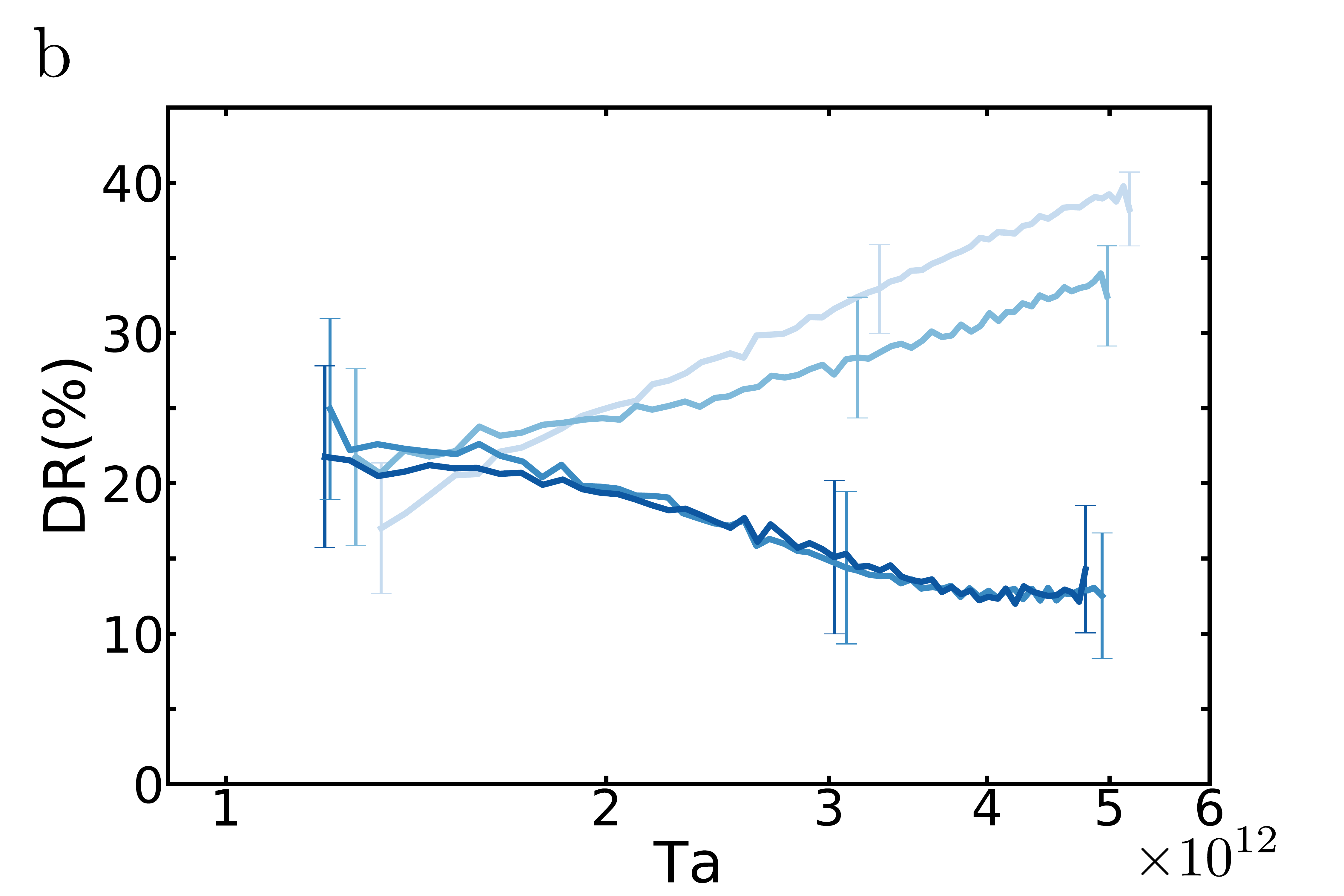}
    \caption{Torque measurements performed in the T$^3$C setup. Black line indicates the reference measurement with no air nor salt, blue lines indicate measurements with an air volume fraction of $\alpha=4\%$. Shade of the coloured lines indicate the salinity, where the light blue line is the decalcified water case and darker lines have a higher salinity. a: Angular velocity Nusselt number. Black dashed line represents the effective scaling $\text{Nu}_\omega\propto\text{Ta}^{0.4}$. b: Drag reduction compared to the reference case, calculated according to equation \ref{eq:Results_DR}.}
    \label{fig:Results_NuTa_DRTa}
\end{figure}

We measure the drag experienced by the flow for salinities in the range $0\leq\Tilde{S}\leq1$, see  figure \ref{fig:Results_NuTa_DRTa}. The black line shows the response of the system in the reference case, where there is no air nor salt ($\Tilde{S}=0$) in the system. The blue lines represent the measurements with an air volume fraction of $\alpha = 4\%$. Darker lines indicate a higher sodium chloride concentration.\\
Figure \ref{fig:Results_NuTa_DRTa}a shows the measured Nusselt number. For the reference case without bubbles and without salts, we observe an effective scaling of $\text{Nu}\propto\text{Ta}^{0.4}$, indicated by the black dashed line, as observed before \cite{vGils_2011,vGils_2012,Huisman_2012,ezeta_2018}. Air present in the system leads to reduction of the Nusselt number for all salt concentrations measured. The case of $\Tilde{S}=0$ has the lowest Nusselt number, i.e., the lowest drag. With increasing salinity the Nusselt number increases, so the angular velocity transport and thus the drag increases. Above $\Tilde{S}=0.75$ no significant difference in the Nusselt number is observed when further increasing the salinity. This could suggest that there is a critical concentration above which increasing the salinity has no additional effect.\\
The drag reduction is calculated comparing the cases with bubbles to the reference case, i.e., using decalcified water as working fluid with no bubbles nor salt. All other measurements are compared to this data. The drag reduction (DR) is calculated as:
\begin{align}\label{eq:Results_DR}
    \begin{split}
        \text{DR}(\text{Ta}) = \frac{\text{Nu}_\omega(\text{Ta},\alpha=0\%) - \text{Nu}_\omega(\text{Ta},\alpha=4\%)}{\text{Nu}_\omega(\text{Ta},\alpha=0\%)},
    \end{split}
\end{align}
see figure \ref{fig:Results_NuTa_DRTa}b. The highest drag reduction is reached for fresh water, i.e., $\Tilde{S}=0$, where we observe drag reduction up to 40\%, as previously reported \cite{Verschoof_2016_surfactant}. The drag reduction becomes less effective with increasing salinity. No clear difference is observed when increasing the salinity above $\Tilde{S}=0.75$. For $\Tilde{S}\leq 0.5$, the drag reduction increases with increasing Taylor number, whereas for $\Tilde{S}\geq0.75$, the drag reduction decreases with increasing Taylor number. We therefore find that in the current Taylor number regime there exist a critical salinity $0.5\leq\Tilde{S}_\text{crit}\leq0.75$ for which we switch from increasing drag reduction to decreasing drag reduction for increasing Ta.\\

\subsection{Estimates of the Weber number}\label{sec:Results_images}
\begin{figure}[!ht]
    \centering
    \includegraphics[width=0.49\textwidth]{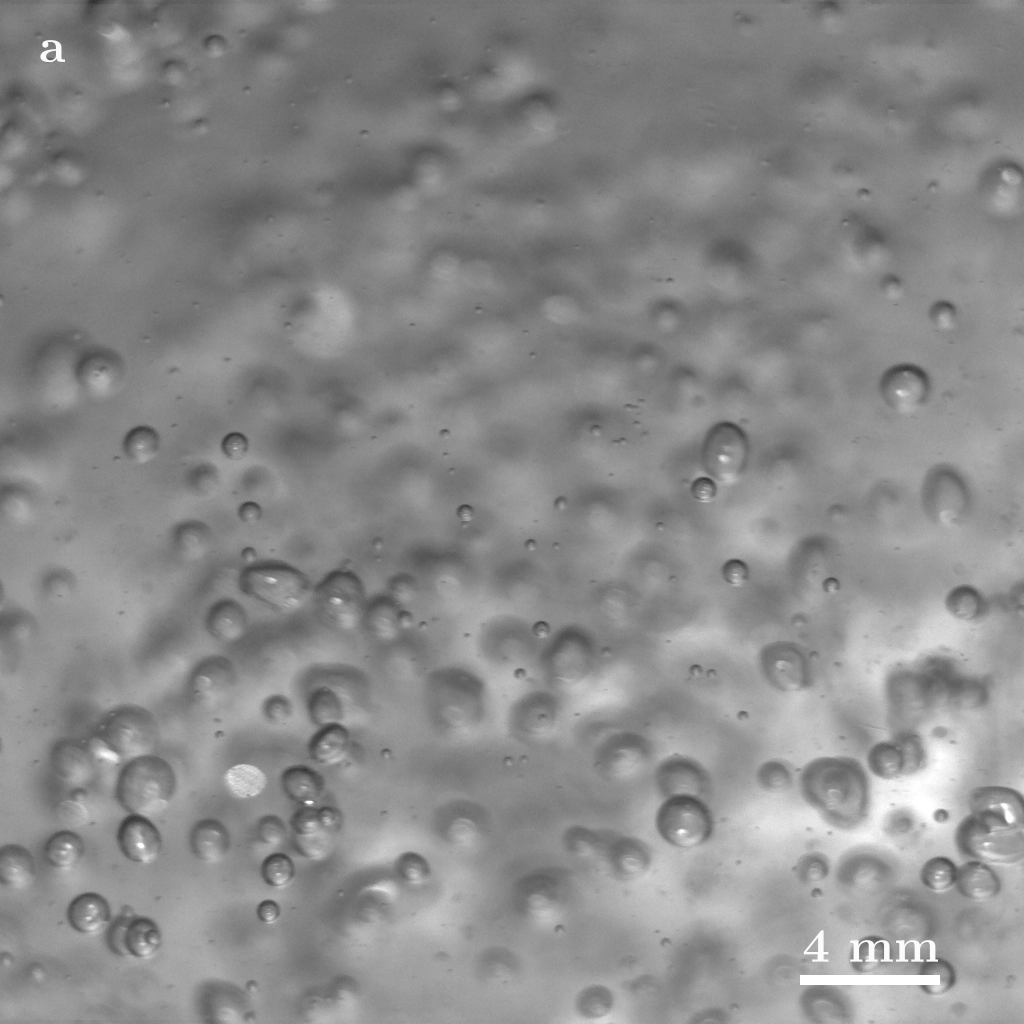}
    \includegraphics[width=0.49\textwidth]{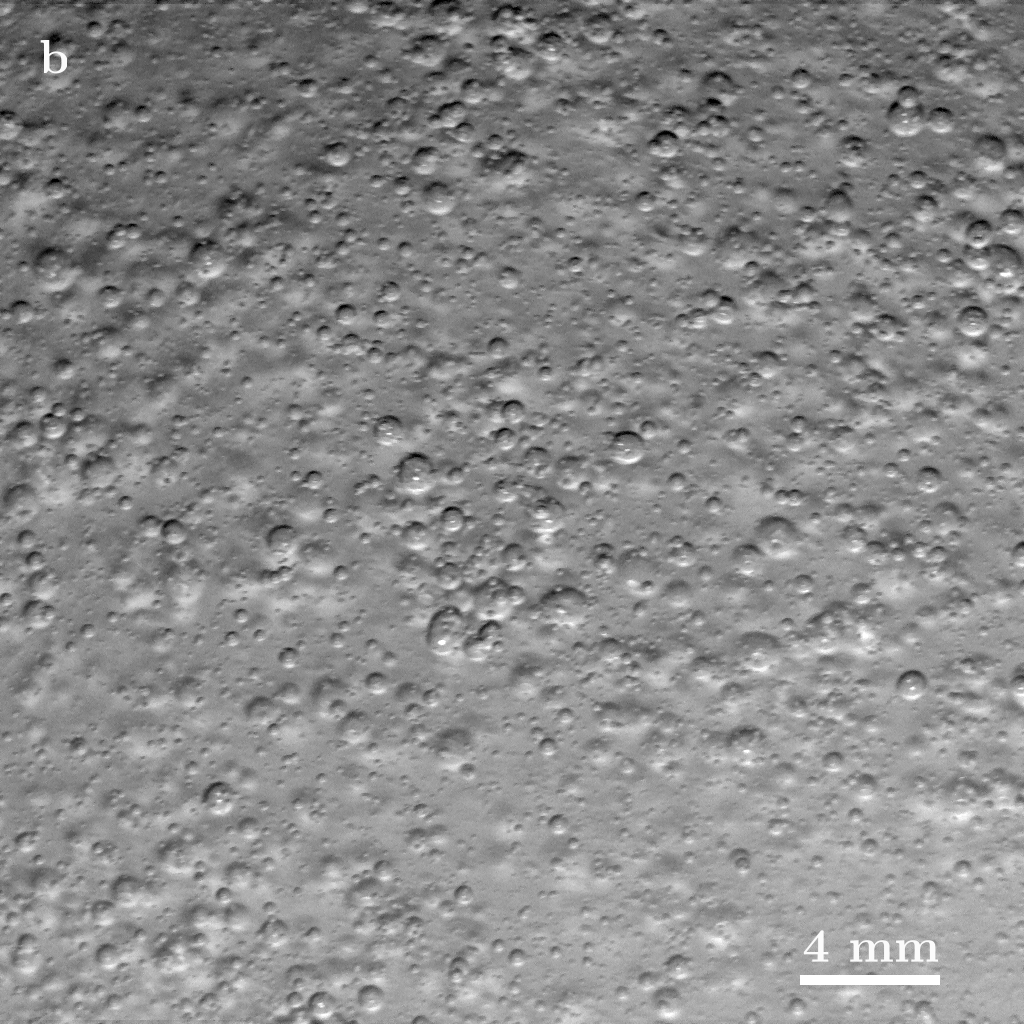}\\
    \includegraphics[width=0.49\textwidth]{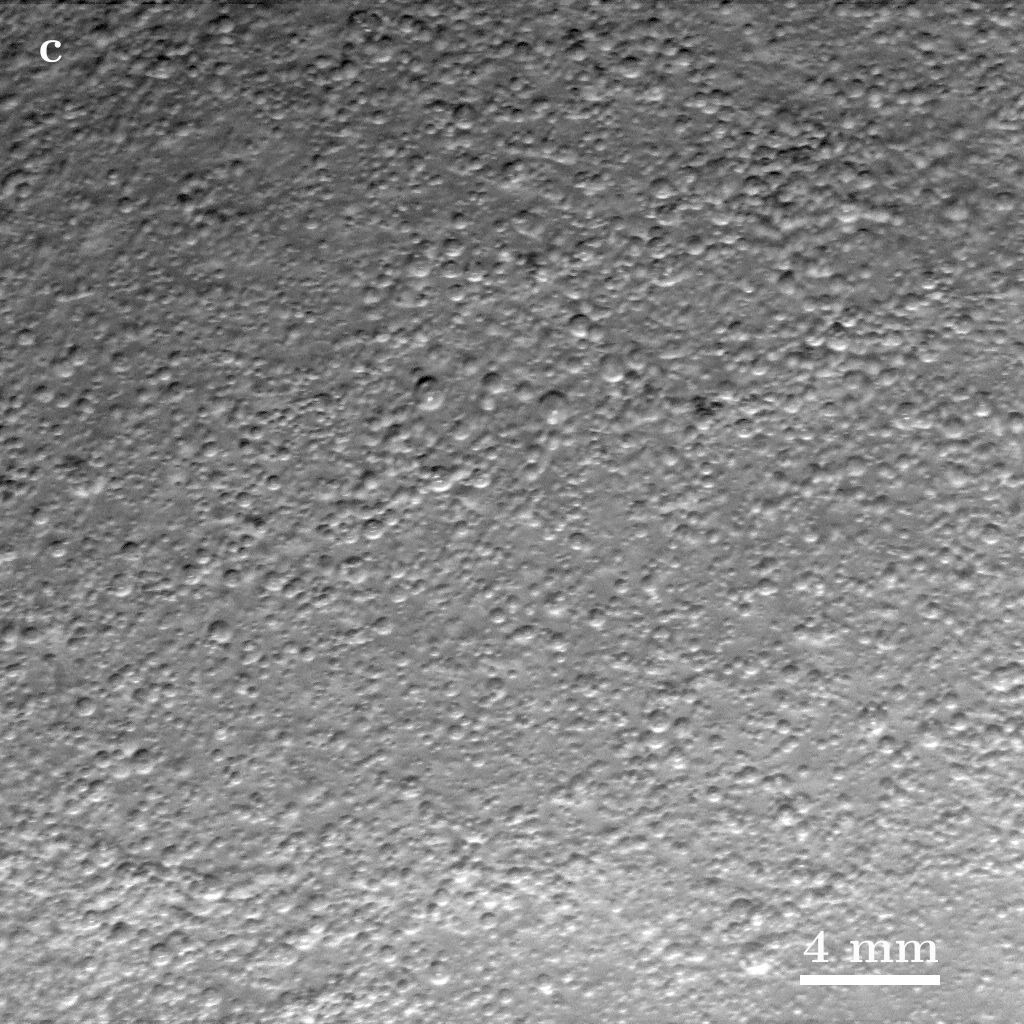}
    \includegraphics[width=0.49\textwidth]{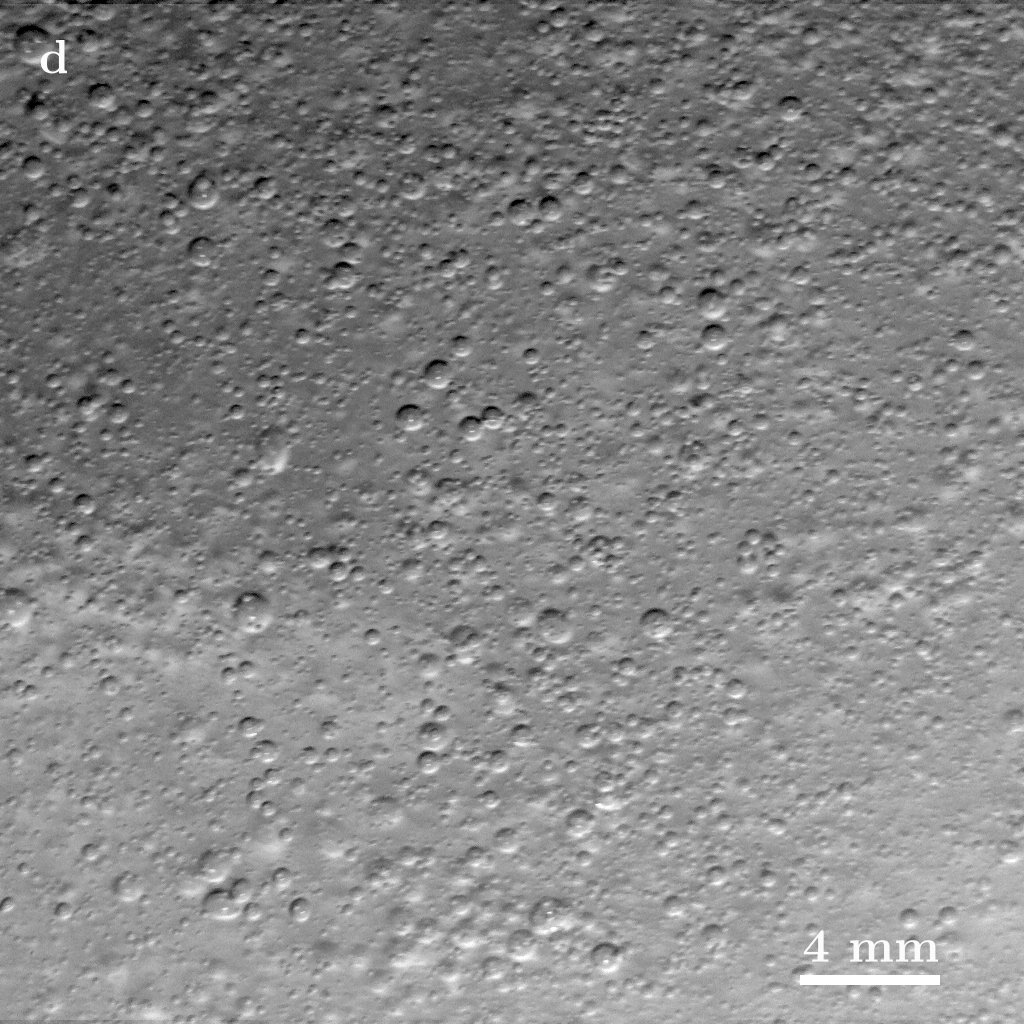}
    \caption{Representative images during measurements with a: $\Tilde{S}=0$, b: $\Tilde{S}=0.5$, c: $\Tilde{S}=0.75$, and d: $\Tilde{S}=1$. Images captured at $Re\approx10^6$ and a global air volume fraction of $\alpha=4\%$ at mid-height, $z=L/2$, and close to the outer cylinder, $r_\text{focal}\approx\unit{0.275}{\meter}$. Field of view is approximately $\unit{30}{\milli\meter}\times\unit{30}{\milli\meter}$. Corresponding videos are included in the supplementary materials.}
    \label{fig:Results_images}
\end{figure}

\begin{figure}[!h]
    \centering
    \includegraphics[width=0.49\textwidth]{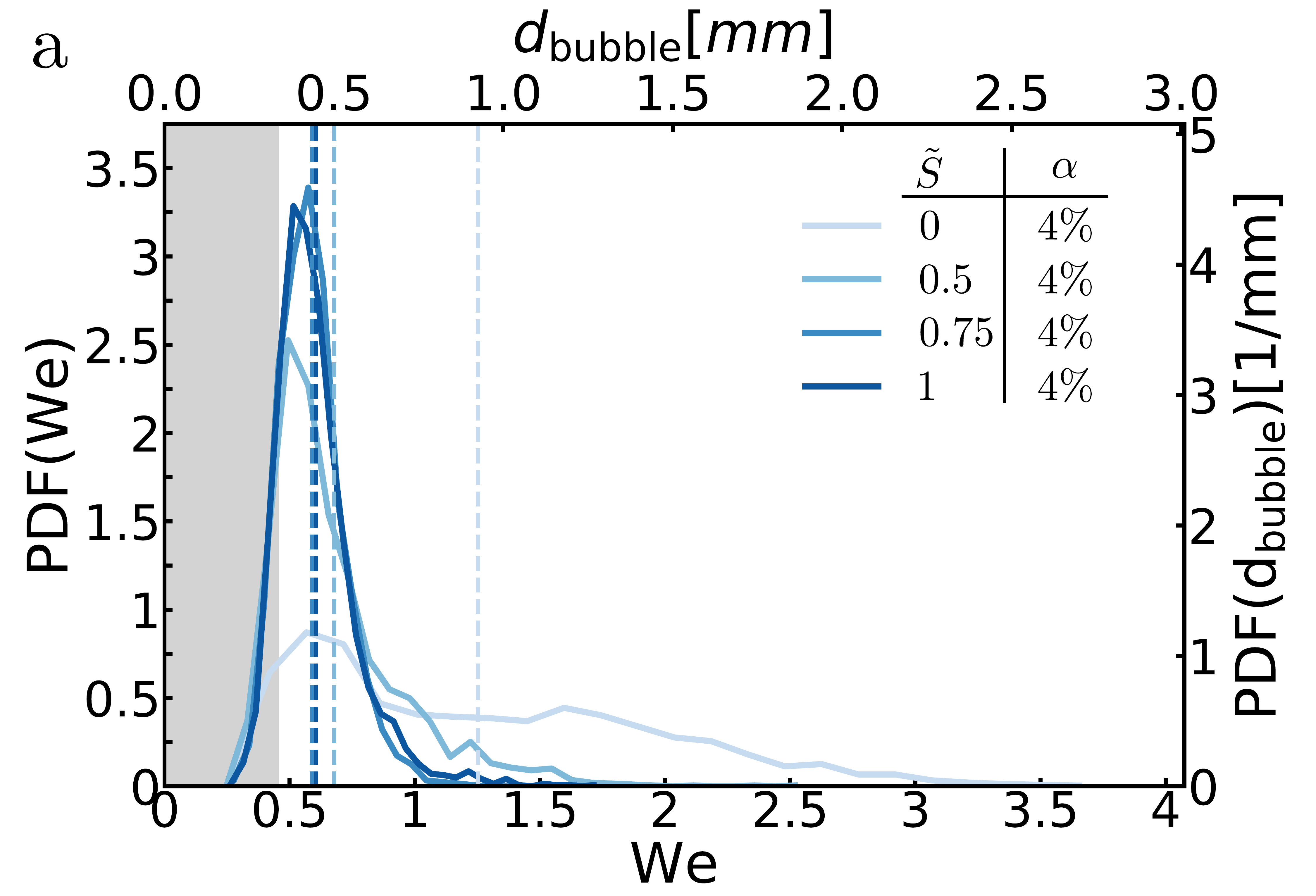}
    \includegraphics[width=0.49\textwidth]{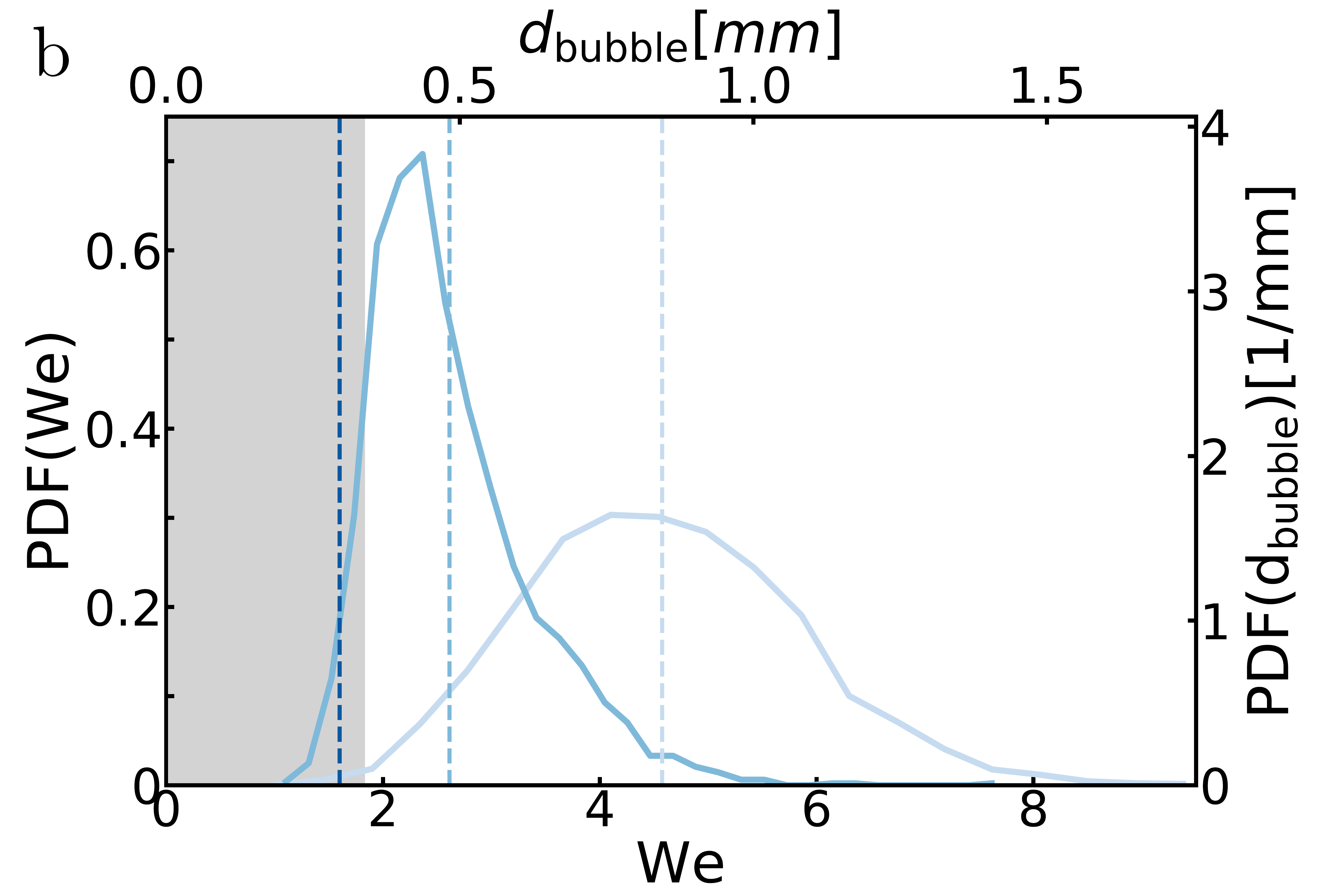}
    \caption{Probability distribution of the Weber number ($\text{We}=\rho u^{\prime2}d_{\text{bubble}}/\gamma$) of the bubbles in the flow for a: $\text{Re}\approx10^6$ ($\eta_K = \unit{18}{\micro\meter}$) b: $\text{Re}\approx2\times10^6$ ($\eta_K = \unit{10}{\micro\meter}$). Colours indicate the salinity and are the same as in figure \ref{fig:Results_NuTa_DRTa}. Vertical dashed lines indicate the mean Weber number for the corresponding measurement. All measurements for a global air volume fraction of $\alpha=4\%$. Bubbles captured at midheight, $z=L/2$, and close to the outer cylinder, $r_\text{focal}\approx\unit{0.275}{\meter}$. Grey areas indicate the region where the bubbles are too small to be (accurately) detected. Top axes indicate the size in\unit{}{\milli\meter} for the case of fresh water and the corresponding PDF axes are on the right. Note that for the cases with salt the conversion is slightly different (by $1.1\%$) due to a different conversion factor $\rho/\gamma$, see table \ref{tab:fluidprop}.}
    \label{fig:Results_bubble_PDF}
\end{figure}

We capture images of the bubbles in the flow using a high speed camera. We focus on bubbles close to the outer cylinder at $r\approx\unit{0.275}{\meter}$, since direct optical access close to the inner cylinder is limited because of the bubbles in the flow. The images are captured at a low frame rate, \unit{50}{fp\second}, as to obtain uncorrelated images and uncorrelated size distributions.\\
Typical images captured during the experiment are shown in figure \ref{fig:Results_images}. Images are captured at $\text{Re}\approx10^6$ (shown in figure \ref{fig:Results_images}) and $\text{Re}\approx2\times10^6$ (not shown). Increasing the salinity leads to smaller bubble sizes in the flow, due to the inhibition of coalescence. The larger bubbles, when $\Tilde{S}=0$, move in bubble clouds and are therefore not homogeneously distributed throughout the volume.\\
The bubble size is directly connected to the Weber number
\begin{align}
\text{We}=\frac{\rho u^{\prime2}d_{\text{bubble}}}{\gamma},
\label{eq:weber}
\end{align}
where $\rho$ is the fluid density, $u^\prime$ are the velocity fluctuations (the standard deviation of $u$), $d_\text{bubble}$ is the diameter of the bubble, and $\gamma$ is the surface tension. For salt water, the presence of the ions will slightly modify the surface tension, and we use the model of ref.\ \cite{Dutcher_2010_surfacetension} to find the surface tension of the salty water, see table \ref{tab:fluidprop}. The smaller the size of the bubble, and thus the smaller their Weber number, means that surface tension becomes more dominant over the forces the bubble experiences in the flow. Therefore the bubble will deform less. The reduced size of the bubbles also decreases their Stokes number and thus increases their mobility, allowing for better mixing inside the flow. The bubbles will therefore be distributed throughout the measurement volume more uniformly.\\
We obtain the velocity fluctuations, which are needed to calculate the Weber number, from the measurements of ref.\ \cite{vGils_2013_bubbledeform}. These authors measured the velocity fluctuations for a global air volume fraction of 3\% in the same setup as we use for our experiments. The velocity fluctuations at the position of the focal plane of the camera are estimated at 2.5\% of the velocity of the inner cylinder.\\
We obtain the probability distribution of the Weber numbers of the bubbles captured by the camera, see figure \ref{fig:Results_bubble_PDF}. The top and right axes represent the bubble size distribution calculated with the density and surface tension of the decalcified water case. The axes are slightly off for the cases with salt, due to a slightly different conversion factor $\rho/\gamma$ in equation \ref{eq:weber} for salt water, see table \ref{tab:fluidprop}. The grey areas indicate when bubbles are too small to be (accurately) detected. The vertical dashed lines indicate the average value of the Weber number for the corresponding measurements. At $\text{Re}\approx10^6$, figure \ref{fig:Results_bubble_PDF}a, the probability of the Weber number of all salinity cases peaks around $\text{We}\approx0.6$. The range of Weber numbers is the widest at $\Tilde{S}=0$. At higher salinities the bubbles become more monodisperse. The long tail at high Weber numbers seen in the PDF for decalcified water disappears when salt is added to the system.\\
At $\text{Re}\approx2\times10^6$, figure \ref{fig:Results_bubble_PDF}b, again we observe the highest Weber numbers in the fresh water case. For $\Tilde{S}\geq0.75$ most bubbles are too small to be accurately detected. Therefore we do not obtain an accurate bubble size distribution and the corresponding Weber number distribution. The images for $\Tilde{S}=0.75$ and $\Tilde{S}=1$ seem to have very similar bubble size distribution when inspected by eye. We estimate that the mean bubble size is $d_\text{bubble}\approx\unit{0.3}{\milli\meter}$, corresponding to a Weber number of $\text{We}\approx1.6$, as indicated by the corresponding vertical dashed line in figure \ref{fig:Results_bubble_PDF}b.\\
We find that the size of the bubbles and their corresponding Weber numbers is not further affected when increasing the salinity above $\Tilde{S}=0.75$. This would suggest that the coalescence is not further inhibited by the presence of extra salts. The resulting critical coalescence concentration ($\Tilde{S}_\text{CCC}$) would be $0.5\leq\Tilde{S}_\text{CCC}\leq0.75$. This is the same range in which we found $\Tilde{S}_\text{crit}$, corresponding to the salt concentration at the transition of increasing drag reduction to decreasing drag reduction with increasing Ta. These two critical concentrations are deducted from two different observations, hence we will treat these as two separate critical concentrations. It is also unclear whether the critical concentrations will always lie within the same range of salinity. The range in which we find the critical coalescence concentration $\Tilde{S}_\text{CCC}$ agrees with the critical coalescence concentration found in ref.\ \cite{Quinn_2014} for bubbles with sizes in the same order of magnitude in a flotation cell. \\
We compare the measured drag reduction with the Weber numbers observed at $\text{Re}\approx2\times10^6$. The highest drag reduction is observed for the fresh water measurements, where also the highest bubble Weber numbers are observed. Increasing the salinity decreases the Weber numbers of the bubbles and correspondingly the drag reduction decreases. When the salinity is increased above $\Tilde{S}=0.75$ the bubble sizes and their corresponding Weber numbers are not further affected. This coincides with the observation that the measured drag reduction is not affected by an increase of salinity above $\Tilde{S}=0.75$. These observations confirm that the bubble size and deformability is of key importance for effective bubbly drag reduction.\\
The near-wall void fraction is often considered to be an important parameter for bubbly drag reduction (see e.g.\ refs.\ \cite{Ceccio_2010_review,Moriguchi_2002}). Although we have not measured the local void fraction, we can obtain these in the case of decalcified water from the measurements of ref.\ \cite{vGils_2013_bubbledeform}. From these measurements we see that the local void fraction near the inner cylinder is higher than in the bulk of the flow. For the cases with higher salinity, the bubbles are smaller and are therefore less affected by the centripetal forces and are more easily transported by the turbulence. It is likely these bubbles are more evenly distributed throughout the measurement volume. We can not exclude that a possible change in distribution of the bubbles also plays a role in the observed change in drag reduction as we are unable to alter the Weber number of the bubbles without changing the Stokes number as well. Numerical investigations could offer more insight in this regard.\\
Comparison with the results from ref.\ \cite{Verschoof_2016_surfactant} indicates that sodium chloride has a less strong effect on the observed Weber numbers and drag reduction than surfactants. While sodium chloride is not as effective as surfactants in decreasing the drag reduction, it is still a non-negligible effect.\\

\section{Conclusion}\label{sec:Conclusions}
We have experimentally investigated the effect of salt on bubbly drag reduction in Taylor--Couette turbulence. Adding sodium chloride to the system, the salinity of the working fluid is increased from fresh water to the average salinity of ocean water. We perform torque measurements combined with high-speed imaging to relate the change in drag reduction to the size of the bubbles in the flow.\\
Whereas in fresh water for a global air volume fraction of 4\% a drag reduction of up to 40\% is achieved, the drag reduction becomes less effective for increasing salinity. We observe a critical salinity, $0.5\leq\Tilde{S}_\text{crit}\leq0.75$, where we switch from increasing drag reduction to decreasing drag reduction with increasing $\text{Ta}$. Above $\Tilde{S}=0.75$, increasing the salinity has no further effect on the measured torque and drag reduction.\\
From the high-speed recordings we obtain the bubble size distribution and calculate the Weber numbers of the captured bubbles. Salts inhibit the coalescence, therefore the bubbles observed in flows containing salts are smaller and have lower Weber numbers. Above $\Tilde{S}=0.75$, an increase in salinity has no further significant effect on the bubble sizes observed in the flow. This indicates a critical coalescence concentration in the range $0.5\leq\Tilde{S}_\text{CCC}\leq0.75$. This range agrees with the critical coalescence concentration of sodium chloride observed in other experiments \cite{Quinn_2014}.\\
We relate the Weber number of the bubbles, characterizing their ability to deform, to the measured drag reduction. Flows with high Weber number bubbles have a high drag reduction, whereas flows with lower Weber number bubbles have a less effective drag reduction. For salinities $\Tilde{S}\geq0.75$, an increase in salinity has no significant effect on the measured drag reduction and Weber numbers of the bubbles in the flow. These findings confirm the importance of bubble size and deformability for effective bubbly drag reduction \cite{vGils_2013_bubbledeform,Verschoof_2016_surfactant}.\\
Our new findings show that considering the mineral content of water is important when investigating bubbly drag reduction for marine applications. In future investigations we want to look into the effect of other salts present in ocean water on bubbly drag reduction. We want to characterize the effects of the salts individually and all of them combined to mimic seawater more closely.\\

\vskip6pt

\enlargethispage{20pt}


\dataccess{Data of figure \ref{fig:Results_NuTa_DRTa} and  figure \ref{fig:Results_bubble_PDF} are added as additional data.}

\aucontribute{LB carried out the experiments, performed the analysis, and drafted the manuscript. DL and SH conceived and designed the study. All authors read, contributed to, and approved the manuscript.}

\competing{The authors declare that they have no competing interests.}

\funding{This work was supported by NWO through the AQUA project.}

\ack{We would like to thank Chao Sun and Yoan Lee for stimulating discussions, Gert-Wim Bruggert, Martin Bos, Thomas Zijlstra, Geert Mentink, and Rindert Nauta for continuous technical support. We thank Vahideh Elhami and Thimo te Molder for performing the ion chromatography measurements. We would like to thank Nouryon for supplying the sodium chloride used for the experiments.}




\begin{thebibliography}{10}

\bibitem{larsson_2010_ship}
Larsson L, Raven HC, Paulling JR.
\newblock {Ship Resistance and Flow}.
\newblock Principles of naval architecture. Society of Naval Architects and
  Marine Engineers; 2010.

\bibitem{Lackenby_1962_ship}
Lackenby H.
\newblock {The Thirty-Fourth Thomas Lowe Gray Lecture: Resistance of Ships,
  with Special Reference to Skin Friction and Hull Surface Condition}.
\newblock Proc Inst Mech Eng. 1962;176(1):981-1014.
\newblock doi:10.1243/PIME\_PROC\_1962\_176\_077\_02.

\bibitem{elbing_2008}
Elbing BR, Winkel ES, Lay KA, Ceccio SL, Dowling DR, Perlin M.
\newblock {Bubble-induced skin-friction drag reduction and the abrupt
  transition to air-layer drag reduction}.
\newblock J Fluid Mech. 2008;612:201-36.
\newblock doi:10.1017/S0022112008003029.

\bibitem{elbing_2013}
Elbing BR, Mäkiharju S, Wiggins A, Perlin M, Dowling DR, Ceccio SL.
\newblock On the scaling of air layer drag reduction.
\newblock J Fluid Mech. 2013;717:484–513.
\newblock doi:10.1017/jfm.2012.588.

\bibitem{Verschoof_2018_cavities}
Verschoof RA, Bakhuis D, Bullee PA, Huisman SG, Sun C, Lohse D.
\newblock {Air cavities at the inner cylinder of turbulent {Taylor–Couette}
  flow}.
\newblock Int J Multiph Flow. 2018;105:264-73.
\newblock doi:10.1016/j.ijmultiphaseflow.2018.04.016.

\bibitem{Zverkhovskyi_2014}
Zverkhovskyi O.
\newblock {Ship Drag Reduction by Air Cavities}.
\newblock TU Delft; 2014.

\bibitem{Ceccio_2010_review}
Ceccio SL.
\newblock {Friction drag reduction of external flows with bubble and gas
  injection}.
\newblock Annu Rev Fluid Mech. 2010;42:183-203.
\newblock doi:10.1146/annurev-fluid-121108-145504.

\bibitem{Murai_2014_review}
Murai Y.
\newblock {Frictional drag reduction by bubble injection}.
\newblock Exp Fluids. 2014;55(7).
\newblock doi:10.1007/s00348-014-1773-x.

\bibitem{Hashim_2015_review}
Hashim A, Yaakob OB, Koh KK, Ismail N, Ahmed YM.
\newblock {Review of micro-bubble ship resistance reduction methods and the
  mechanisms that affect the skin friction on drag reduction from 1999 to2015}.
\newblock J Teknol. 2015;74(5):105-14.

\bibitem{lohse_2018}
Lohse D.
\newblock {Bubble puzzles: From fundamentals to applications}.
\newblock Phys Rev Fluids. 2018 Nov;3:110504.
\newblock doi:10.1103/PhysRevFluids.3.110504.

\bibitem{Madavan_1984}
Madavan NK, Deutsch S, Merkle CL.
\newblock {Reduction of turbulent skin friction by microbubbles}.
\newblock Phys Fluids. 1984;27(2):356-63.
\newblock doi:10.1063/1.864620.

\bibitem{Madavan_1985}
Madavan NK, Deutsch S, Merkle CL.
\newblock {Measurements of local skin friction in a microbubble-modified
  turbulent boundary layer}.
\newblock J Fluid Mech. 1985;156:237-56.
\newblock doi:10.1017/S0022112085002075.

\bibitem{xu_2002}
Xu J, Maxey MR, Karniadakis GEM.
\newblock {Numerical simulation of turbulent drag reduction using
  micro-bubbles}.
\newblock J Fluid Mech. 2002;468:271–281.
\newblock doi:10.1017/S0022112002001659.

\bibitem{Moriguchi_2002}
Moriguchi Y, Kato H.
\newblock Influence of microbubble diameter and distribution on frictional
  resistance reduction.
\newblock J Mar Sci Technol. 2002;7(2):79-85.
\newblock doi:10.1007/s007730200015.

\bibitem{Lo_2006}
Lo TS, L'vov V, Procaccia I.
\newblock {Drag reduction by compressible bubbles}.
\newblock Phys Rev E. 2006 04;73:036308.
\newblock doi:10.1103/PhysRevE.73.036308.

\bibitem{kato_1999}
Kato H, Iwashina T, Miyanaga M, Yamaguchi H.
\newblock {Effect of microbubbles on the structure of turbulence in a turbulent
  boundary layer}.
\newblock J Mar Sci Technol. 1999 12;4:155-62.
\newblock doi:10.1007/PL00010624.

\bibitem{ferrante_elghobashi_2004}
Ferrante A, Elghobashi S.
\newblock On the physical mechanisms of drag reduction in a spatially
  developing turbulent boundary layer laden with microbubbles.
\newblock J Fluid Mech. 2004;503:345–355.
\newblock doi:10.1017/S0022112004007943.

\bibitem{vdBerg_2005}
van~den Berg TH, Luther S, Lathrop DP, Lohse D.
\newblock {Drag Reduction in Bubbly Taylor-Couette Turbulence}.
\newblock Phys Rev Lett. 2005 Jan;94:044501.
\newblock doi:10.1103/PhysRevLett.94.044501.

\bibitem{vGils_2013_bubbledeform}
van Gils DPM, Narezo~Guzman D, Sun C, Lohse D.
\newblock {The importance of bubble deformability for strong drag reduction in
  bubbly turbulent Taylor–Couette flow}.
\newblock J Fluid Mech. 2013 5;722:317-47.
\newblock doi:10.1017/jfm.2013.96.

\bibitem{Kitagawa_2005_bubbledeform}
Kitagawa A, Hishida K, Kodama Y.
\newblock {Flow structure of microbubble-laden turbulent channel flow measured
  by PIV combined with the shadow image technique}.
\newblock Exp Fluids. 2005;38(4):466-75.
\newblock doi:10.1007/s00348-004-0926-8.

\bibitem{Lu_2005_bubbledeform}
Lu J, Fern{\'a}ndez A, Tryggvason G.
\newblock {The effect of bubbles on the wall drag in a turbulent channel flow}.
\newblock Phys Fluids. 2005;17(9):1-12.
\newblock doi:10.1063/1.2033547.

\bibitem{Verschoof_2016_surfactant}
Verschoof RA, van~der Veen RCA, Sun C, Lohse D.
\newblock {Bubble Drag Reduction Requires Large Bubbles}.
\newblock Phys Rev Lett. 2016 Sep;117:104502.
\newblock doi:10.1103/PhysRevLett.117.104502.

\bibitem{Murai_2005}
Murai Y, Oiwa H, Takeda Y.
\newblock Bubble behavior in a vertical Taylor-Couette flow.
\newblock J Phys Conf Ser. 2005 jan;14:143-56.
\newblock doi:10.1088/1742-6596/14/1/018.

\bibitem{murai_2008}
Murai Y, Oiwa H, Takeda Y.
\newblock {Frictional drag reduction in bubbly Couette–Taylor
  flow}.
\newblock Phys Fluids. 2008;20(3):034101.
\newblock doi:10.1063/1.2884471.

\bibitem{sugiyama_2008}
Sugiyama K, Calzavarini E, Lohse D.
\newblock Microbubbly drag reduction in Taylor–Couette flow in the wavy
  vortex regime.
\newblock J Fluid Mech. 2008;608:21–41.
\newblock doi:10.1017/S0022112008001183.

\bibitem{spandan_2018}
Spandan V, Verzicco R, Lohse D.
\newblock Physical mechanisms governing drag reduction in turbulent
  Taylor–Couette flow with finite-size deformable bubbles.
\newblock J Fluid Mech. 2018;849:R3.
\newblock doi:10.1017/jfm.2018.478.

\bibitem{silverstream_site}
Technologies S. {What is air lubrication?};.
\newblock Accessed: 15 July 2022.
\newblock Available from:
  \url{https://www.silverstream-tech.com/what-is-air-lubrication/}.

\bibitem{Mizokami_2010_mitsubishi}
Mizokami S, Kawakita C, Kodan Y, Takano S, Higasa S, Shigenaga R.
\newblock {Experimental Study of Air Lubrication Method and Verification of
  Effects on Actual Hull by Means of Sea Trial}.
\newblock Mitsubishi Heavy Industries Technical Review. 2010 Sep;47.

\bibitem{Mizokami_2013_mitsubishi}
Mizokami S, Kawakado M, Kawano M, Hasegawa T, Hirakawa I.
\newblock {Implementation of Ship Energy--Saving Operations with Mitsubishi Air
  Lubrication System}.
\newblock Mitsubishi Heavy Industries Technical Review. 2013 Jun;50.

\bibitem{Kumagai_2015}
Kumagai I, Takahashi Y, Murai Y.
\newblock {Power-saving device for air bubble generation using a hydrofoil to
  reduce ship drag: Theory, experiments, and application to ships}.
\newblock Ocean Eng. 2015;95:183-94.
\newblock doi:10.1016/j.oceaneng.2014.11.019.

\bibitem{Kodama_2000}
Kodama Y, Kakugawa A, Takahashi T, Kawashima H.
\newblock {Experimental study on microbubbles and their applicability to ships
  for skin friction reduction}.
\newblock Int J Heat Fluid Flow. 2000;21(5):582-8.
\newblock doi:10.1016/S0142-727X(00)00048-5.

\bibitem{Jang_2014}
Jang J, Choi SH, Ahn SM, Kim B, Seo JS.
\newblock {Experimental investigation of frictional resistance reduction with
  air layer on the hull bottom of a ship}.
\newblock Int J Nav Archit Ocean Eng. 2014;6(2):363-79.
\newblock doi:10.2478/IJNAOE-2013-0185.

\bibitem{vdBerg_2007}
van~den Berg TH, van Gils DPM, Lathrop DP, Lohse D.
\newblock Bubbly Turbulent Drag Reduction Is a Boundary Layer Effect.
\newblock Phys Rev Lett. 2007 Feb;98:084501.
\newblock doi:10.1103/PhysRevLett.98.084501.

\bibitem{Bullee_2020_hydrophobic}
Bullee PA, Verschoof RA, Bakhuis D, Huisman SG, Sun C, Lammertink RGH, Lohse D.
\newblock {Bubbly drag reduction using a hydrophobic inner cylinder in
  Taylor–Couette turbulence}.
\newblock J Fluid Mech. 2020;883:A61.
\newblock doi:10.1017/jfm.2019.894.

\bibitem{Bullee_2020_roughness}
Bullee PA, Bakhuis D, Ezeta R, Huisman SG, Sun C, Lammertink RGH, Lohse D.
\newblock {{Effect of axially varying sandpaper roughness on bubbly drag
  reduction in Taylor–Couette turbulence}}.
\newblock Int J Multiph Flow. 2020:103434.
\newblock doi:https://doi.org/10.1016/j.ijmultiphaseflow.2020.103434.

\bibitem{Deutsch_2004}
Deutsch S, Moeny M, Fontaine A, Petrie H.
\newblock Microbubble drag reduction in rough walled turbulent boundary layers
  with comparison against polymer drag reduction.
\newblock Exp Fluids. 2004 11;37:731-44.
\newblock doi:10.1007/s00348-004-0863-6.

\bibitem{vanBuren_2017}
Van~Buren T, Smits AJ.
\newblock {Substantial drag reduction in turbulent flow using liquid-infused
  surfaces}.
\newblock J Fluid Mech. 2017;827:448–456.
\newblock doi:10.1017/jfm.2017.503.

\bibitem{ezeta_2019}
Ezeta R, Bakhuis D, Huisman SG, Sun C, Lohse D.
\newblock Drag reduction in boiling Taylor–Couette turbulence.
\newblock J Fluid Mech. 2019;881:104–118.
\newblock doi:10.1017/jfm.2019.758.

\bibitem{gillissen_2013}
Gillissen JJJ.
\newblock Turbulent drag reduction using fluid spheres.
\newblock J Fluid Mech. 2013;716:83–95.
\newblock doi:10.1017/jfm.2012.510.

\bibitem{sanders_2006}
Sanders WC, Winkel ES, Dowling DR, Perlin M, Ceccio SL.
\newblock Bubble friction drag reduction in a high-Reynolds-number flat-plate
  turbulent boundary layer.
\newblock J Fluid Mech. 2006;552:353–380.
\newblock doi:10.1017/S0022112006008688.

\bibitem{murai_2007}
Murai Y, Fukuda H, Oishi Y, Kodama Y, Yamamoto F.
\newblock Skin friction reduction by large air bubbles in a horizontal channel
  flow.
\newblock Int J Multiph Flow. 2007;33(2):147-63.
\newblock doi:https://doi.org/10.1016/j.ijmultiphaseflow.2006.08.008.

\bibitem{Winkel_2004}
Winkel ES, Ceccio SL, Dowling DR, Perlin M.
\newblock {Bubble-size distributions produced by wall injection of
  air into flowing freshwater, saltwater and surfactant solutions}.
\newblock Exp Fluids. 2004;37:802-10.
\newblock doi:10.1007/s00348-004-0850-y.

\bibitem{shen_2006}
Shen X, Ceccio SL, Perlin M.
\newblock {Influence of bubble size on micro-bubble drag reduction}.
\newblock Exp Fluids. 2006;41(3):415-24.
\newblock doi:10.1007/s00348-006-0169-y.

\bibitem{Craig_1993}
Craig VSJ, Ninham BW, Pashley RM.
\newblock {Effect of electrolytes on bubble coalescence}.
\newblock Nature. 1993;364(6435):317-9.
\newblock doi:10.1038/364317a0.

\bibitem{Hori_2020}
Hori Y, Bothe D, Hayashi K, Hosokawa S, Tomiyama A.
\newblock {Mass transfer from single carbon-dioxide bubbles in
  surfactant-electrolyte mixed aqueous solutions in vertical pipes}.
\newblock Int J Multiph Flow. 2020;124:103207.
\newblock doi:https://doi.org/10.1016/j.ijmultiphaseflow.2020.103207.

\bibitem{Firouzi_2014}
Firouzi M, Nguyen AV.
\newblock {Effects of monovalent anions and cations on drainage and lifetime of
  foam films at different interface approach speeds}.
\newblock Adv Powder Technol. 2014;25(4):1212-9.
\newblock doi:https://doi.org/10.1016/j.apt.2014.06.004.

\bibitem{Firouzi_2015}
Firouzi M, Howes T, Nguyen AV.
\newblock {A quantitative review of the transition salt concentration for
  inhibiting bubble coalescence}.
\newblock Adv Colloid Interface Sci. 2015;222:305-18.
\newblock Reinhard Miller, Honorary Issue.
\newblock doi:https://doi.org/10.1016/j.cis.2014.07.005.

\bibitem{Quinn_2014}
Quinn JJ, Sovechles JM, Finch JA, Waters KE.
\newblock {Critical coalescence concentration of inorganic salt solutions}.
\newblock Miner Eng. 2014;58:1-6.
\newblock doi:https://doi.org/10.1016/j.mineng.2013.12.021.

\bibitem{Sovechles_2015}
Sovechles JM, Waters KE.
\newblock {Effect of ionic strength on bubble coalescence in inorganic salt and
  seawater solutions}.
\newblock AIChE J. 2015;61(8):2489-96.
\newblock doi:10.1002/aic.14851.

\bibitem{Tsang_2004}
Tsang YH, Koh YH, Koch DL.
\newblock {Bubble-size dependence of the critical electrolyte concentration for
  inhibition of coalescence}.
\newblock J Colloid Interface Sci. 2004;275(1):290-7.
\newblock doi:https://doi.org/10.1016/j.jcis.2004.01.026.

\bibitem{vGils_2011_T3C}
van Gils DPM, Bruggert GW, Lathrop DP, Sun C, Lohse D.
\newblock {{The Twente turbulent Taylor–Couette (T3C) facility: Strongly
  turbulent (multiphase) flow between two independently rotating cylinders}}.
\newblock Rev Sci Instrum. 2011;82(2):.
\newblock doi:http://dx.doi.org/10.1063/1.3548924.

\bibitem{Taylor_1923}
Taylor GI.
\newblock {Stability of a viscous liquid contained between two rotating
  cylinders}.
\newblock Philos Trans R Soc A. 1923;223:289-343.
\newblock doi:10.1098/rsta.1923.0008.

\bibitem{andereck_1986}
Andereck CD, Liu SS, Swinney HL.
\newblock Flow regimes in a circular Couette system with independently rotating
  cylinders.
\newblock J Fluid Mech. 1986;164:155–183.
\newblock doi:10.1017/S0022112086002513.

\bibitem{Grossmann_2016_TCrev}
Grossmann S, Lohse D, Sun C.
\newblock {High--Reynolds Number Taylor-Couette Turbulence}.
\newblock Annu Rev Fluid Mech. 2016;48(1):53-80.
\newblock doi:10.1146/annurev-fluid-122414-034353.

\bibitem{ostillamonico_2014}
Ostilla-Mónico R, van~der Poel EP, Verzicco R, Grossmann S, Lohse D.
\newblock Exploring the phase diagram of fully turbulent Taylor–Couette flow.
\newblock J Fluid Mech. 2014;761:1–26.
\newblock doi:10.1017/jfm.2014.618.

\bibitem{Fardin_2014}
Fardin MA, Perge C, Taberlet N.
\newblock ``The hydrogen atom of fluid dynamics'' -- introduction to the
  Taylor--Couette flow for soft matter scientists.
\newblock Soft Matter. 2014;10:3523-35.
\newblock doi:10.1039/C3SM52828F.

\bibitem{Eckhardt_2007}
Eckhardt B, Grossmann S, Lohse D.
\newblock {Torque scaling in turbulent Taylor–Couette flow between
  independently rotating cylinders}.
\newblock J Fluid Mech. 2007 6;581:221-50.
\newblock doi:10.1017/S0022112007005629.

\bibitem{ahlers_2009}
Ahlers G, Grossmann S, Lohse D.
\newblock {Heat transfer and large scale dynamics in turbulent Rayleigh-Bénard
  convection}.
\newblock Rev Mod Phys. 2009 Apr;81:503-37.
\newblock doi:10.1103/RevModPhys.81.503.

\bibitem{einstein_1905}
Einstein A.
\newblock {Eine neue bestimmung der molek{\"u}ldimensionen}.
\newblock ETH Zurich; 1905.

\bibitem{Chouippe_2014}
Chouippe A, Climent E, Legendre D, Gabillet C.
\newblock Numerical simulation of bubble dispersion in turbulent Taylor-Couette
  flow.
\newblock Phys Fluids. 2014;26(4):043304.
\newblock doi:10.1063/1.4871728.

\bibitem{Climent_2007}
Climent E, Simonnet M, Magnaudet J.
\newblock Preferential accumulation of bubbles in Couette-Taylor flow patterns.
\newblock Phys Fluids. 2007;19(8):083301.
\newblock doi:10.1063/1.2752839.

\bibitem{Djeridi_2004}
Djeridi H, Gabillet C, Billard JY.
\newblock Two-phase Couette--Taylor flow: Arrangement of the dispersed phase
  and effects on the flow structures.
\newblock Phys Fluids. 2004;16(1):128-39.
\newblock doi:10.1063/1.1630323.

\bibitem{Mehel_2005}
Mehel A, Gabillet C, Djeridi H.
\newblock {Bubble Effect on the Structures of Weakly Turbulent Couette Taylor
  Flow}.
\newblock Fluids Eng Division Summer Meeting. 2005 06;Volume 1: Symposia, Parts
  A and B:751-60.
\newblock doi:10.1115/FEDSM2005-77020.

\bibitem{bakhuis_2021}
Bakhuis D, Ezeta R, Bullee PA, Marin A, Lohse D, Sun C, Huisman SG.
\newblock Catastrophic Phase Inversion in High-Reynolds-Number Turbulent
  Taylor-Couette Flow.
\newblock Phys Rev Lett. 2021 Feb;126:064501.
\newblock doi:10.1103/PhysRevLett.126.064501.

\bibitem{Fokoua_2015}
Fokoua GN, Gabillet C, Aubert A, Colin C.
\newblock Effect of bubble's arrangement on the viscous torque in bubbly
  Taylor-Couette flow.
\newblock Phys Fluids. 2015;27(3):034105.
\newblock doi:10.1063/1.4915071.

\bibitem{Huisman_2012}
Huisman SG, van Gils DPM, Grossmann S, Sun C, Lohse D.
\newblock {Ultimate Turbulent Taylor-Couette Flow}.
\newblock Phys Rev Lett. 2012 Jan;108:024501.
\newblock doi:10.1103/PhysRevLett.108.024501.

\bibitem{Pereira_2001_dens}
Pereira G, Moreira R, Vázquez MJ, Chenlo F.
\newblock {Kinematic viscosity prediction for aqueous solutions with
  various solutes}.
\newblock Chem Eng J. 2001;81(1):35-40.
\newblock doi:https://doi.org/10.1016/S1385-8947(00)00203-5.

\bibitem{Aleksandrov_2012_visc}
Aleksandrov AA, Dzhuraeva EV, Utenkov VF.
\newblock {Viscosity of aqueous solutions of sodium chloride}.
\newblock High Temp. 2012;50(3):354-8.
\newblock doi:10.1134/S0018151X12030029.

\bibitem{Dutcher_2010_surfacetension}
Dutcher CS, Wexler AS, Clegg SL.
\newblock {Surface Tensions of Inorganic Multicomponent Aqueous
  Electrolyte Solutions and Melts}.
\newblock J Phys Chem A. 2010;114(46):12216-30.
\newblock doi:10.1021/jp105191z.

\bibitem{vGils_2011}
van Gils DPM, Huisman SG, Bruggert GW, Sun C, Lohse D.
\newblock {Torque Scaling in Turbulent Taylor-Couette Flow with Co- and
  Counterrotating Cylinders}.
\newblock Phys Rev Lett. 2011 Jan;106:024502.
\newblock doi:10.1103/PhysRevLett.106.024502.

\bibitem{vGils_2012}
van Gils DPM, Huisman SG, Grossmann S, Sun C, Lohse D.
\newblock {Optimal Taylor–Couette turbulence}.
\newblock J Fluid Mech. 2012;706:118–149.
\newblock doi:10.1017/jfm.2012.236.

\bibitem{ezeta_2018}
Ezeta R, Huisman SG, Sun C, Lohse D.
\newblock {Turbulence strength in ultimate Taylor–Couette turbulence}.
\newblock J Fluid Mech. 2018;836:397–412.
\newblock doi:10.1017/jfm.2017.795.

\end{thebibliography}

\end{document}